\documentclass[11pt,sort&compress]{elsarticle}
\makeatletter
\def\ps@pprintTitle{%
 \let\@oddhead\@empty
 \let\@evenhead\@empty
 \def\@oddfoot{\centerline{\thepage}}%
 \let\@evenfoot\@oddfoot}
\makeatother
\usepackage{amsmath}
\usepackage{accents}
\usepackage{amssymb}
\usepackage{mathrsfs}
\usepackage[LGR,T1]{fontenc}
\usepackage[latin1]{inputenc}
\usepackage[cal=boondox,calscaled=1]{mathalfa}
\DeclareMathAlphabet{\bbvar}{U}{BOONDOX-ds}{m}{n}
\DeclareMathAlphabet{\bbgreek}{U}{bbold}{m}{n}
\usepackage[defaultsans,osfigures,scale=0.9]{opensans}
\usepackage{graphicx}
\usepackage{psfrag}
\usepackage{pstool}
\usepackage{caption}
\usepackage{graphicx}%
\newcommand{\hook}{\text{\large{$\lrcorner$}}}
\usepackage{color}
\usepackage{hyperref}
\newcommand{\qq}[1]{``#1''} %Anfuehrungszeichen
\newcommand{\utilde}[1]{\underaccent{\tilde}{#1}}
\newcommand{\di}{\mathrm{d}}
\usepackage{slashed}
\usepackage{tensor}\newcommand{\ou}[3]{{#1}{}^{#2}{}_{#3}}
\newcommand{\uo}[3]{{#1}{}_{#2}{}^{#3}}

\newcommand{\uepsilon}{{\underaccent{\tilde}{\epsilon}}}
\newcommand{\oepsilon}{{\tilde{\epsilon}}}

\newcommand{\I}{\mathrm{i}} %imaginaere Einheit
\newcommand{\E}{\mathrm{e}} %Euler Zahl
\newcommand{\CC}{\mathrm{cc.}} % komplex konjugiertes
\newcommand{\HC}{\mathrm{hc.}} % hermitsch konjugiertes
\newcommand{\C}{\mathbb{C}}

\newcommand{\R}{\mathbb{R}}

\newenvironment{subalign}{\subequations\align}{\endalign\endsubequations}
\newcommand{\eref}[1]{(\ref{#1})}

\DeclareMathAlphabet{\sfit}{OT1}{fos}{sb}{it}
\DeclareMathAlphabet{\mathsf}{OT1}{fos}{sb}{n}
\begin{document}

\begin{abstract}
In this paper, I investigate the quantisation of length in euclidean quantum gravity in three dimensions. The starting point is the classical hamiltonian formalism in a cylinder of finite radius. At this finite boundary, a counter term is introduced that couples the gravitational field in the interior to a two-dimensional conformal field theory for an $SU(2)$ boundary spinor, whose norm determines the conformal factor between the fiducial boundary metric and the physical metric in the bulk. The equations of motion for this boundary spinor are derived from the boundary action and turn out to be the two-dimensional analogue of the Witten equations appearing in Witten's proof of the positive mass theorem. The paper concludes with some comments on the resulting quantum theory. It is shown, in particular, that the length of a one-dimensional cross section of the boundary turns into a number operator on the Fock space of the theory. The spectrum of this operator is discrete and matches the results from loop quantum gravity in the spin network representation.% This is an important consistency check for the program.
\end{abstract}
%%%%%%\tile{Spinors as boundary variables in classical and quantum general relativity}
\title{Quantum gravity in three dimensions, Witten spinors and the quantisation of length}
%\title{Klassische und quantenmechanische Freiheitsgrade der Gravitation auf lichartigen Rändern}
\author{Wolfgang Wieland}
\address{Perimeter Institute for Theoretical Physics\\31 Caroline Street North\\ Waterloo, ON N2L\,2Y5, Canada\\{\vspace{0.5em}\normalfont Fall 2017}
}
\date{Fall 2017}
%\keywords{test}
\maketitle
{\vspace{-1.2em}
{\tableofcontents}
%:
\begin{center}{\noindent\rule{\linewidth}{0.4pt}}\end{center}
\section{Introduction}
\noindent %Loop quantum gravity has demonstrated that a theory of quantum theory may indeed be a quantum theory of geometry itself. In four spacetime dimensions, there are operators for length, area, angles and volume, and all of these operators have a discrete spectrum. %The construction of these operators relies on the so-called spin-network representation: The fundamental excitations of geometry are generated from superpositions of products of gravitational Wilson lines, and each of these Wilson lines represents a fundamental quantum of area. 
One of the key open issues for loop quantum gravity is to check (or prove it impossible) that the fundamental quantum discreteness of space that we see in the theory is compatible with the known physics in the continuum. The question is, in other words, how to go from a theory with only finitely many degrees of freedom on a spin network graph to a field theory with infinitely many propagating degrees of freedom. In this paper, I will turn this question around, and show that in three dimensions the loop gravity quantisation of space can be understood already from the theory in the continuum without ever introducing spin networks or triangulations of space. An analogous argument for Lorentzian gravity in four dimensions appeared in the previous paper \cite{Wieland:2017cmf} in this series.

Now, in three dimensions, gravity is topological, and there are no local degrees of freedom in the bulk. The situation becomes more interesting if boundaries are included. Boundaries typically break gauge invariance (such as diffeomorphism invariance) and what was an unphysical pure gauge direction before may now turn into an actual physical degree of freedom at the boundary. At infinity, the dynamics of such boundary modes for three-dimensional gravity is typically governed by a two-dimensional conformal field theory \cite{Brown:1986nw, Witten:1998qj,carlipbook, Carlip:2005zn}. The question is then if such a construction exists at finite distance as well. The goal of this paper is to demonstrate that such a boundary field theory exists and can be constructed in terms of an $SU(2)$ boundary spinor coupled to the gravitational field in the bulk. The choice for spinors as boundary variables for pure gravity may seem a little odd, 
but it fits well into the picture that we get from non-perturbative quantum gravity, where the fundamental excitations of geometry (in three spacetime dimensions) are given by gravitational Wilson lines for an $SU(2)$ spin connection. If these gravitational Wilson lines hit a boundary they excite a surface charge, namely an $SU(2)$ spinor sitting at the puncture. The purpose of this paper is to investigate the field theory for such boundary spinors in the continuum. 
%\footnote{References \cite{Dittrich:2017hnl,Dittrich:2017rvb} are themselves part of a wider effort to realise non-perturbative quantum gravity in finite regions, see also \cite{Donnelly:2016auv,Geiller:2017xad} and \cite{wieland:nulldefects,Wieland:2017cmf,Wieland:2017zkf}.}

Two recent developments in the field support this idea: first of all, a pair of papers \cite{Dittrich:2017hnl,Dittrich:2017rvb} have appeared quite recently that studied the boundary theory of the Ponzano\,--\,Regge spinfoam model \cite{ponzanoregge,Freidel:1998pt,alexreview}. The authors evaluate the Ponzano\,--\,Regge spinfoam amplitudes $Z_{\mathrm{PR}}[\cdot]$ against boundary coherent states \cite{Livine:2007vk} at the \emph{finite} boundary of a solid torus. These coherent states $\Psi_{\underline{\xi}}$ are labelled by spinors $(\xi^A_1,\xi^A_2,\dots)$ that saturate the open legs of spin networks stretching into the bulk (see \hyperref[fig1]{figure 1}). The evaluation of the amplitudes for such boundary states defines then an effective boundary action $\E^{\I S_{\mathrm{eff}}[\underline{\xi}]}\sim Z_{\mathrm{PR}}[\Psi_{\underline{\xi}}]$, whose critical points define a classical lattice model for the boundary spinors. That such a theory should exist then also at finite boundaries in the continuum is motivated by another development in the field: during the last couple of years a new representation was developed for four-dimensional\footnote{In three dimensions, such a representation exists as well, and the Ponzano\,--\,Regge amplitudes can be derived, in fact, from a one-dimensional worldline model \cite{Wieland:2014ab} for such $SU(2)$ boundary spinors alone.} loop quantum gravity in terms of  $SL(2,\C)$ spinors \cite{komplexspinors,twistintegrals,wieland:nulldefects, Wieland:2017zkf,Wieland:2017cmf}. At the level of classical general relativity these spinors can be understood as gravitational boundary variables on a null surface: the canonical pair consists of a surface spinor (the \emph{null flag} of the boundary) and a conjugate spinor-valued two-form \cite{Wieland:2017zkf}.\vspace{0.4em} %In addition, there are also information theoretical arguments that support the idea to build quantum gravity from quantising  from the qfundamental variables for quantum gravity in finite regions \cite{Mueller:2016aov}.\vspace{0.4em}  %The generalisation to four dimensions is obviously more difficult. So far the following results have appeared in the literature: %,Bianchi:2016hmk

The paper is divided into two parts. The first part develops the classical field theory for the boundary spinors and investigates the equations of motion and their relation to the Witten equation. The second part deals with the hamiltonian formulation of the field equations and the gauge symmetries of the theory, which are internal $SU(2)$ frame rotations and small diffeomorphisms. In fact, only those small diffeomorphisms that vanish at the boundary are genuine gauge transformations of the theory. There are then also those large diffeomorphisms $\varphi:\mathcal{M}\rightarrow\mathcal{M}$ that do not vanish at the boundary but map it onto itself: $\varphi(\partial{\mathcal{M}})=\partial{\mathcal{M}}$. Indeed, these are genuine symmetries of the theory, and there is an infinite number of them (because there are infinitely many diffeomorphisms that preserve the boundary). Finally, we discuss some aspects of the resulting quantum theory, in particular the quantisation of the conformal factor, which is given by the norm of the boundary spinor. On the physical phase space, the components of this boundary spinor satisfy the commutation relations of the harmonic oscillator. In quantum theory, the conformal factor turns therefore into the number operator on the Fock space of the theory. The physical length of a one-dimensional cross section of the boundary is determined by the possible eigenvalues of this operator. The resulting spectrum is discrete and agrees with the results from loop quantum gravity in three dimensions \cite{Rovelli:1993kc}. The entire derivation happens at the level of the continuum theory, and no spin networks or triangulations of space are ever required for deriving this result. %The starting point of the construction is the derivation of the symplectic structure on the physical phase space at finite distance. Gravity in the bulk is topological, therefore the only contribution to the symplectic structure comes from the boundary. At this boundary canonical variables are identified that satisfy the usual harmonic oscillator Poisson brackets. The number operator for these oscillators turns out to have a geometric meaning\,--\,it defines the conformal factor between the physical metric in the bulk and the diucial and two-dimensional metric at the boundary. The metrical length of teh boundary with respect to the p The entire derivation happens at the level of the continuum theory, and no spin-networks or SU(2) gauge variables are ever required for deriving this result.

\begin{figure}[h]
\begin{center}
\psfrag{A}{$\mathcal{C}$}
\psfrag{B}{$\mathcal{B}$}
\psfrag{C}{$\mathcal{M}$}
\psfrag{D}{${\Sigma}$}
\psfrag{E}{$\mathcal{C}$}
\psfrag{F}{{\footnotesize$\xi^A$}}
\psfrag{G}{{\footnotesize$\Psi$}}
\hspace{1.5em}\includegraphics[width=0.7\textwidth]{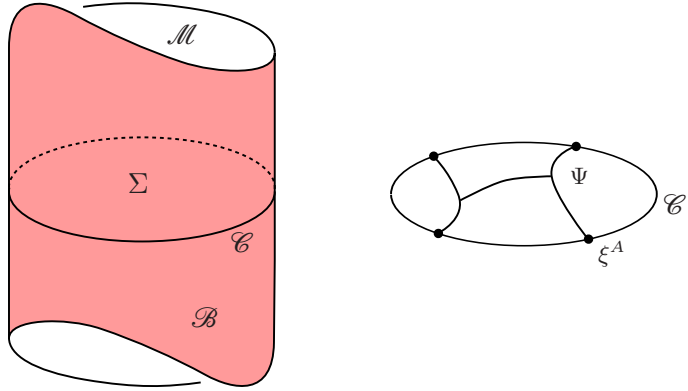}
\end{center}
\caption{\emph{Left:} We study three-dimensional euclidean gravity in an infinitely tall cylinder $\mathcal{M}\simeq\R\times\Sigma$. Its boundary is the two-dimensional world-tube $\mathcal{B}\simeq\R\times S^1$. The hypersurface ${\Sigma}$ intersects this boundary in a circular line $\mathcal{C}\simeq S^1$, which is assumed to have a \emph{finite} length. \emph{Right:} In three-dimensional loop quantum gravity, the quantum states $\Psi$ of geometry are constructed from two-dimensional (planar) spin networks, which are built from gravitational Wilson lines. These Wilson lines may hit the boundary, where they create a spinor-valued surface operator $\xi^A$. The purpose of the paper is to study these \emph{loop gravity boundary spinors} from the perspective of the classical field theory.}\label{fig1}
\end{figure}
\section{Action and equations of motion}
\subsection{Action and boundary terms}
\noindent In the absence of a cosmological constant, the vacuum Einstein equations follow from the topological $BF$ action\footnote{We are using units of $\hbar=c=1$, and we are in three dimensions, hence Newton's constant $G$ has dimensions of $\text{length}\sim\text{mass}^{-1}$. }
\begin{equation}
S_{\mathcal{M}}[e,A]=-\frac{1}{8\pi G}\int_{\mathcal{M}}e_i\wedge F^i[A].\label{bulkactn}
\end{equation}
The action is a functional of the $SU(2)$ spin connection $\ou{A}{i}{a}$ and the frame fields $\ou{e}{i}{a}$ that diagonalise the metric tensor
\begin{equation}
g_{ab}=\delta_{ij}\ou{e}{i}{a}\ou{e}{j}{b},\label{three-metrc}
\end{equation}
where $\delta_{ij}$ denotes the flat and internal Euclidean metric (internal indices $i,j,k,\dots$ are raised and lowered using this metric). The resulting equations of motion are the flatness constraint
\begin{equation}
F^i=\di A^i+\frac{1}{2}\ou{\epsilon}{i}{jk}A^j\wedge A^k=0,\label{fltnss}
\end{equation}
and the torsionless condition
\begin{equation}
T^i=\nabla e^i=\di e^i+\ou{\epsilon}{i}{jk}A^j\wedge e^k,\label{tors}
\end{equation}
where $\ou{\epsilon}{i}{jk}$ are the $SU(2)$ structure constants and $\nabla$ is the gauge covariant exterior derivative. The torsionless condition determines the spin rotation coefficients $\ou{A}{i}{a}$ uniquely\footnote{The connection is unique provided that $\ou{e}{i}{a}$ has an inverse, i.e.\ $\exists\uo{e}{i}{a}:\uo{e}{i}{a}\ou{e}{i}{b}=\delta^a_b$.} and we are left therefore with a locally flat metric manifold $(\mathcal{M},g_{ab})$ in the bulk. At the boundary, on the other hand, the variation of the connection yields the remainder
\begin{equation}
\delta_AS_{\mathcal{M}}[e,A]\approx\frac{1}{8\pi G}\int_{\partial\mathcal{M}}e_i\wedge\delta A^i,\label{remindr}
\end{equation}
where \qq{$\approx$} denotes equality up to terms that vanish provided the equations of motion are satisfied.

The goal is then to introduce a boundary field theory, whose action will compensate the boundary term \eref{remindr} coming from the bulk. This theory should be both $SU(2)$ gauge invariant and local. The integral \eref{remindr} is linear in the connection, and we are thus looking for a boundary action that is linear in the connection as well. The most minimal fields that the $SU(2)$ gauge covariant derivative can act upon are spinors. This motivates us to consider a two-dimensional Dirac action for an $SU(2)$ boundary spinor $\xi^A$, which is minimally coupled to the gauge connection $\ou{A}{i}{a}$. We will consider therefore the boundary field theory defined by the following action
\begin{equation}
S_{\partial\mathcal{M}}[\xi,q|A]=\frac{1}{2\I}\int_{\partial\mathcal{M}}\Big(\xi^\dagger_A\ou{\sigma}{A}{Bi}q^i\wedge D\xi^B-\CC\Big),\label{bndryactn}
\end{equation}
where $\ou{\sigma}{A}{Bi}$ are the Pauli matrices (the relevant conventions are explained in the \hyperref[appndx]{appendix}) and  $D_a$ is the pull-back of the three-dimensional covariant derivative to the boundary: if $\varphi_{\partial\mathcal{M}}:\partial\mathcal{M}\hookrightarrow\mathcal{M}$ denotes the canonical embedding,
\begin{equation}
D_a:=\varphi_{\partial\mathcal{M}}^\ast \nabla_a.\label{Ddef}
\end{equation}
We have introduced here an additional frame field at the boundary, namely $\ou{q}{i}{a}$, which is an $\mathfrak{su}(2)$ Lie algebra-valued one-form intrinsic to the boundary. The corresponding two-dimensional boundary metric is given by
\begin{equation}
q_{ab}=\delta_{ij}\ou{q}{i}{a}\ou{q}{j}{b}.
\end{equation}
Indices intrinsic to the boundary will be raised and lowered using $q_{ab}$ and its inverse: $q^{ab}q_{bc}=\ou{q}{a}{c}$. 

The frame fields $\ou{q}{i}{a}$ define a linear map $V_i\mapsto V_j\ou{q}{j}{a}$ from the three-dimensional space of Euclidean three-vectors into $T^\ast(\partial\mathcal{M})$, which is two-dimensional. Hence there is one degenerate direction, which we call
\begin{equation}
n^i:n_i\ou{q}{i}{a}=0,\,n_in^i=1.\label{bndryn}
\end{equation}
We can define then also the two-dimensional Levi-Civita tensors
\begin{equation}
\varepsilon_{ij}:=n^m\epsilon_{mij},\quad\text{and}\quad\varepsilon_{ab}=\varepsilon_{ij}\ou{q}{i}{a}\ou{q}{j}{b}.
\end{equation}
It will be also useful to define the following vector-valued boundary one-form, which will play the role of the extrinsic curvature, namely
\begin{equation}
\ou{K}{i}{a}=D_an^i.\label{exdef}
\end{equation}

\subsection{Glueing conditions}
\noindent The bulk plus boundary theory is defined now by the action
\begin{equation}
S_q[e,A|\xi]=-\frac{1}{8\pi G}\int_{\mathcal{M}}e_i\wedge F^i[A]+\frac{1}{2\I}\int_{\mathcal{B}}\Big(\xi^\dagger_A\ou{\sigma}{A}{Bi}q^i\wedge D\xi^B-\CC\Big).
\end{equation}
The boundary frame fields $\ou{q}{i}{a}$ are external background fields which are held fixed in the variational principle (modulo $SU(2)$ gauge transformations $\ou{q}{i}{a}\rightarrow\ou{\epsilon}{i}{jk}\Lambda^j\ou{q}{k}{a}$, diffeomorphisms and local conformal transformations $\ou{q}{i}{a}\rightarrow\E^\lambda\ou{q}{i}{a}$). 

The equations of motion derived from the variation of the action split then into those defined in the bulk and those propagating the boundary fields along the cylinder $\mathcal{B}=\partial\mathcal{M}$. The variation of the dreibein $\ou{e}{i}{a}$ yields the torsionless condition \eref{tors} in the bulk, the variation of the connection, on the other hand, yields the flatness constraint \eref{fltnss} and a remainder at the boundary,
\begin{equation}
\delta_AS_q[e,A|\xi]\approx\frac{1}{8\pi G}\int_{\mathcal{B}}e_i\wedge\delta A^i-\frac{1}{4}\int_{\mathcal{B}}\Big[\xi^\dagger_A\ou{\sigma}{A}{Ci}\ou{\sigma}{C}{Bj}\xi^B+\CC\Big]q^i\wedge\delta A^j.
\end{equation}
Using the Pauli identity \eref{Pauliid}, and setting this variation to zero, we find the following glueing condition,
\begin{equation}
\varphi^\ast_{\mathcal{B}}\ou{e}{i}{a}=4\pi G\,\|\xi\|^2\,\ou{q}{i}{a}.\label{glucond}
\end{equation}
In other words, the pull-back of the triad to the boundary is given by the fiducial boundary triad $\ou{q}{i}{a}$ times a conformal factor, which is proportional to the norm of the $SU(2)$ boundary spinor.

\subsection{Boundary field theory and the Witten equation}
\noindent At the boundary, we now have additional field equations as well. The critical points of the boundary action with respect to variations of $\xi^A$ are given by those field configurations that satisfy
\begin{equation}
\ou{\sigma}{A}{Ba}\varepsilon^{ab}D_b\xi^B-\frac{1}{2}\ou{\sigma}{A}{Bi}\vartheta^i\xi^B=0,\label{bdynmcs}
\end{equation}
where $\ou{\sigma}{A}{Ba}=\ou{\sigma}{A}{Bi}\ou{q}{i}{a}$ are the boundary soldering forms, and $\vartheta^i$ measures the torsion of $D_a$ with respect to the boundary triad $\ou{q}{i}{a}$,
\begin{equation}
\vartheta^i:=\varepsilon^{ab}D_a\ou{q}{i}{b}.\label{torsdef}
\end{equation}
The vanishing of torsion \eref{tors} in the bulk implies that this internal three vector is tangent to the boundary, hence $\vartheta^in_i=0$. 

Let us then write the boundary equations of motion \eref{bdynmcs} in a more geometrical language. We introduce, therefore, the $U(1)$ intrinsic spin connection to the boundary, together with the corresponding covariant derivative $\eth_a$, which has the following properties,
\begin{equation}
\varepsilon^{ab}\eth_{a}\ou{q}{i}{b}=0,\quad
\eth_a q_{bc}=0,\quad \eth_an^i=0.
\end{equation}
The relation between $\eth_a$ and $D_a$ is given by a difference tensor $\ou{\Delta}{i}{a}$, which is defined as follows,
\begin{equation}
(D_a-\eth_a)V^i=\ou{\epsilon}{i}{jk}\ou{\Delta}{j}{a}V^k.
\end{equation}
Going back to the definition for the boundary torsion \eref{torsdef} and the extrinsic curvature \eref{exdef}, we can decompose now the difference tensor into its tangential and normal contributions, namely
\begin{equation}
\ou{\Delta}{i}{a}=-n^i\vartheta_a-\ou{\epsilon}{i}{j}\ou{K}{j}{a},
\end{equation}
where $\vartheta_a=\vartheta_i\ou{q}{i}{a}$. If we now use the fundamental Pauli identity \eref{Pauliid}, we can rewrite the boundary equations of motion into the following compact form,
\begin{equation}
\ou{\sigma}{A}{Ba}\eth^a\xi^B+\frac{1}{2}\ou{n}{A}{B}\ou{K}{a}{a}\xi^B=0.\label{Witteq}
\end{equation}
If the second term vanished, this would just be the ordinary two-dimensional Dirac equation $\slashed{\eth}\xi^A=\ou{\sigma}{A}{Ba}q^{ab}\eth_a\xi^B=0$. Yet this second term does not vanish in general: it is constructed from the trace $\ou{K}{a}{a}=\uo{q}{i}{a}\ou{K}{i}{a}$ of the extrinsic curvature \eref{exdef} and from $\ou{n}{A}{B}:=\ou{\sigma}{A}{Bi}n^i$, which is the matrix-valued internal boundary normal \eref{bndryn}. In general, the second term will contribute therefore non-trivially. The resulting equation is well known in general relativity. It is the two-dimensional analogue of the Witten equation (the Dirac equation at the boundary coupled to the trace of the extrinsic curvature) that Witten used in his celebrated proof of the positive mass theorem \cite{Wittenproof}. Here, the Witten equation emerges as well, but with a very different role\,---\,it defines the very dynamics of the gravitational boundary degrees of freedom.
\section{Boundary observables and quantisation of length}
\subsection{Symplectic structure, gauge symmetries, observables}
\noindent Slicing the cylinder along a hypersurface $\Sigma$ into two halves (see \hyperref[fig1]{figure 1}), we evaluate the first variation of the bulk plus boundary action on-shell, and identify the covariant pre-symplectic structure,
\begin{equation}
\Theta_\Sigma=\frac{1}{8\pi G}\int_\Sigma e_i\wedge\bbvar{d}A^i+\frac{1}{2\I}\int_{\mathcal{C}}\Big[\xi^\dagger_A\ou{\tilde{\sigma}}{A}{B}\bbvar{d}\xi^B-\CC\Big].\label{thetadef}
\end{equation}
The symbol \qq{$\bbvar{d}$} denotes the exterior derivative on the covariant phase space (the space of solutions of the theory), and $\ou{\tilde{\sigma}}{A}{B}$ is the matrix-valued line density\footnote{Introducing a local coordinate $s$ on $\mathcal{C}$, we can write this density simply as the densitised Pauli matrix $\ou{\tilde{\sigma}}{A}{B}=ds\sqrt{q_{ab}\partial^a_s\partial^b_s}\,\ou{\sigma}{A}{Bi}\ou{q}{i}{c}\partial^c_s.$  }
\begin{equation}
\ou{\tilde{\sigma}}{A}{B}=\ou{\sigma}{A}{Bi}\,\varphi^\ast_{\mathcal{C}}q^i,\label{sigmadens}
\end{equation}
with $\varphi^\ast_{\mathcal{C}}:T^\ast\mathcal{B}\rightarrow T^\ast\mathcal{C}$ denoting the pull-back to the one-dimensional cross-section $\mathcal{C}=\partial\Sigma$. It is also useful to introduce the corresponding inverse matrix-valued density $\ou{\utilde{\sigma}}{A}{B}$ on $\mathcal{C}$, which is defined implicitly by
\begin{equation}
\ou{\underaccent{\tilde}{\sigma}}{A}{C}\ou{\tilde{\sigma}}{C}{B}=\delta^A_B.
\end{equation}

The pre-symplectic two-form is given then by the exterior derivative of the symplectic potential,
\begin{equation}
\Omega_\Sigma=\bbvar{d}\Theta_\Sigma=\frac{1}{8\pi G}\int_{\Sigma}\bbvar{d}e_i\wedge\bbvar{d}A^i-\I\int_{\mathcal{C}}\bbvar{d}\xi^\dagger_A\ou{\tilde{\sigma}}{A}{B}\bbvar{d}\xi^B.\label{Omdef}
\end{equation}
Notice, in particular, that the matrix-valued density $\ou{\tilde{\sigma}}{A}{B}$ is taken as an external background structure. On phase space, all field variations of $\ou{\tilde{\sigma}}{A}{B}$ vanish, and the only non-vanishing Poisson brackets (at the pre-symplectic or kinematical level) between the fundamental phase-space variables are therefore given by 
\begin{equation}
\big\{\tensor[^2]{e}{^i_a}(p),\tensor[^2]{A}{^j_b}(p^\prime)\big\}=8\pi G\,\delta^{ij}\uepsilon_{ab}\tilde{\delta}^{(2)}(p,p'),\label{bulkpoiss}
\end{equation}
and
\begin{equation}
\big\{\xi^A(s),\xi^\dagger_B(s')\big\}=-\ou{\utilde{\sigma}}{A}{B}\tilde{\delta}^{(1)}(s,s').\label{xipoiss}
\end{equation}
In here, $\uepsilon_{ab}$ is the inverse Levi-Civita tensor density on $\Sigma$ and $\tensor[^2]{e}{^i_a}$ (resp.\ $\tensor[^2]{A}{^i_a}$) denotes the pull-back of $\ou{e}{i}{a}$ (resp.\ ${}\ou{A}{i}{a}$) to $\Sigma$, while $\tilde{\delta}^{(n)}(\cdot,\cdot)$ is the $n$-dimensional Dirac distribution (a density on resp.\ $\Sigma$ and $\mathcal{C}$). The kinematical phase space is thus coordinatised by a triple of fields $(\tensor[^2]{e}{^i_a},\tensor[^2]{A}{^i_a},\xi^A)$.

For a generic vector field $t^a\in T\mathcal{B}$ and an arbitrary foliation $\{\Sigma_t\}_{t\in\R}$ of the cylinder, a subtlety arises, because the line density $\ou{\tilde{\sigma}}{A}{B}=\varphi^\ast_{\Sigma_t}\ou{\sigma}{A}{B}$ (which defines the symplectic structure) will be now time dependent (i.e.\ $\mathcal{L}_t\ou{\tilde{\sigma}}{A}{B}:=\varphi^\ast_{\Sigma_t}\mathcal{L}_t\ou{\sigma}{A}{B}\neq 0$). The appearance of this explicit $t$-dependence affects the Hamilton equations, which are modified by the introduction of a covariant derivative. This can be seen as follows: consider some general hamiltonian $H_t[\tensor[^2]{e}{^i_a},\tensor[^2]{A}{^i_a},\xi^A]$ on phase space, and the equations of motion derived from the action
\begin{equation}
S=\int_\gamma \di t\big(\Theta_{\Sigma_t}(\delta_t)-H_t\big),
\end{equation}
where $\delta_t$ is the time derivative. In the interior of $\Sigma$, the equations of motion will assume the familiar hamiltonian form,
\begin{equation}
\delta_t[\tensor[^2]{A}{^i_a}]=\big\{H_t,\tensor[^2]{A}{^i_a}\big\},\quad
\delta_t[\tensor[^2]{e}{^i_a}]=\big\{H_t,\tensor[^2]{e}{^i_a}\big\}.\label{covD1}
\end{equation}
At the boundary $\mathcal{C}=\partial\Sigma$, we have to take into account that $\ou{\tilde{\sigma}}{A}{B}$ may be itself  time dependent, hence $\delta_t[\ou{\tilde{\sigma}}{A}{B}]\neq 0$. In deriving the equations of motion from the variation of the action (for boundary conditions $\Theta_{\Sigma_t}(\delta_t)=0$ on $\partial\gamma$) the vector field $\delta_t$ will hit now $\ou{\tilde{\sigma}}{A}{B}$ and modify, therefore, the hamiltonian equations of motion. In fact, only the left hand side of the equations of motion is modified: the ordinary derivative is replaced by the covariant derivative
\begin{equation}
\mathcal{D}_t[\xi^A]:=\delta_t[\xi^A]+\ou{\omega}{A}{B}(\delta_t)\xi^B=\big\{H_t,\xi^A\big\},\label{covD2}
\end{equation}
for a connection $\ou{\omega}{A}{B}$ on phase space, which is given by
\begin{equation}
\ou{\omega}{A}{B}(\delta_t)=\frac{1}{2}\ou{\utilde{\sigma}}{A}{C}\delta_t[\ou{\tilde{\sigma}}{C}{B}]\xi^B,
\end{equation}
where $\delta_t[\ou{\tilde{\sigma}}{A}{B}]$ is inferred from the glueing condition \eref{glucond}.

We can now generalise this idea and say that a general field  variation $\delta_\varepsilon$ on phase space (for some gauge parameter $\varepsilon$) defines a hamiltonian charge $Q_\varepsilon$ if for all other field variations $\delta$ on phase space the integrability condition
\begin{equation}
\Omega_\Sigma(\mathcal{D}_\varepsilon,\delta)=-\delta Q_\varepsilon
\end{equation}
is satisfied. Notice, that we have interpreted here the covariant derivative $\mathcal{D}_\varepsilon$ as a vector field on phase space, whose components (at the level of the covariant phase space) are given by
\begin{subequations}\begin{align}
\mathcal{D}_\varepsilon[\ou{e}{i}{a}]&=\delta_\varepsilon[\ou{e}{i}{a}],\\
\mathcal{D}_\varepsilon[\ou{A}{i}{a}]&=\delta_\varepsilon[\ou{A}{i}{a}],\\
\mathcal{D}_\varepsilon[\xi^A]&=\delta_\varepsilon[\xi^A]+\ou{\omega}{A}{B}(\delta_\varepsilon)\xi^B.
\end{align}%
\label{covD3}%
\end{subequations}%
The gauge symmetries of the theory are given then by the degenerate directions of $\Omega_\Sigma$, and we will see in a moment that in this sense both internal $SU(2)$ frame rotations and bulk diffeomorphisms are gauge symmetries of the theory. On the other hand, finite diffeomorphisms that do not vanish at the cylindrical boundary $\mathcal{B}$ (but map it onto itself) are generated by boundary observables, and there are infinitely many such observables, because there are infinitely many vector fields that preserve the boundary.\vspace{0.4em}

\noindent{\emph{(i) internal gauge transformations.}} First of all, we consider internal frame rotations. At the Lagrangian level they are generated by the vector field $\delta_\Lambda$, whose bulk and boundary components are given by
\begin{equation}
\begin{split}
&\delta_\Lambda\ou{e}{i}{a}=\ou{\epsilon}{i}{lm}\Lambda^l\ou{e}{m}{a},\\
&\delta_\Lambda\ou{q}{i}{a}=\ou{\epsilon}{i}{lm}\Lambda^l\ou{q}{m}{a},\phantom{\frac{1}{2\I}}
\end{split}
\begin{split}
&\delta_\Lambda\ou{A}{i}{a}=-\nabla_a\Lambda^i,\\
&\delta_\Lambda\xi^A=\frac{1}{2\I}\ou{\sigma}{A}{Bi}\Lambda^i\xi^B\equiv\ou{\Lambda}{A}{B}\xi^B,
\end{split}
\end{equation}
for a gauge parameter $\Lambda^i$. On phase space, the corresponding covariant derivative (see \eref{covD1}, \eref{covD2}, \eref{covD3}) is given by
\begin{equation}
\mathcal{D}_\Lambda\xi^A=\delta_\Lambda\xi^A+\ou{\omega}{A}{B}(\delta_\Lambda)\xi^B=\frac{1}{2}\ou{\Lambda}{A}{B}\xi^B+\frac{1}{2}\tensor{\big(\utilde{\sigma}\Lambda\tilde{\sigma}\big)}{^A_B}\xi^B.
\end{equation}

Let then $\delta$ be a second and linearly independent field variation  (a linearised solution of the field equations (\ref{fltnss}, \ref{tors}, \ref{glucond}, \ref{bdynmcs}), hence a tangent vector to the covariant phase space). We  evaluate the pre-symplectic two-form \eref{Omdef} on these vector fields, and obtain
\begin{align}
\nonumber\Omega_\Sigma(\mathcal{D}_\Lambda,\delta)=&\frac{1}{8\pi G}\int_\Sigma\Big(\epsilon_{ilm}\Lambda^le^m\wedge\delta A^i+\delta e_i\wedge\nabla\Lambda^i\Big)+\\
\nonumber&+\frac{1}{2\I}\int_{\mathcal{C}}\Big(-\frac{1}{2}\xi^\dagger_A\ou{\Lambda}{A}{C}\ou{\tilde{\sigma}}{C}{B}\delta\xi^B-\frac{1}{2}\xi^\dagger_A\ou{\tilde\sigma}{A}{C}\ou{\Lambda}{C}{B}\delta\xi^B+\\
\nonumber&\qquad-\frac{1}{2}\delta\xi^\dagger_A\ou{\tilde{\sigma}}{A}{C}\ou{\Lambda}{C}{B}\xi^B-\frac{1}{2}\delta\xi^\dagger_A\ou{\Lambda}{A}{C}\ou{\tilde{\sigma}}{C}{B}\xi^B-\CC\Big)=\\
=&-\frac{1}{8\pi G}\int_{\mathcal{C}}\Lambda_i\delta e^i+\frac{1}{2}\int_{\mathcal{C}}\Lambda^iq_i\delta\|\xi\|^2.
\end{align}
Going from the first to the last line we used Stokes's theorem, the vanishing of torsion in the bulk \eref{tors} and the Pauli identity \eref{Pauliid}. In addition $\ou{q}{i}{a}$ is treated as an external background structure, whose variation vanishes on phase space, i.e.\ $\delta\ou{q}{i}{a}=0$. The glueing conditions \eref{glucond} imply  that the last line vanishes, hence
\begin{equation}
\Omega_\Sigma(\mathcal{D}_\Lambda,\cdot)=0.
\end{equation}
Internal $SU(2)$ frame rotations, including even those large gauge transformations that do not vanish at the boundary, are therefore gauge symmetries of the theory.\vspace{0.4em}

%\noindent \emph{(ii) local conformal transformations.} A short moment of reflection shows that the local conformal transformations 
%\begin{equation}
%%\delta_\lambda\xi^A=\frac{\lambda}{2}\xi^A,\quad\delta_\lambda\ou{q}{i}{a}=-\lambda\ou{q}{i}{a},
%\end{equation}
%for  local gauge parameters $\lambda:\mathcal{C}\rightarrow\R$ define degenerate directions of the pre-symplectic two-form, hence
%\begin{equation}
%\Omega_\Sigma(\delta_\lambda,\cdot)=0.
%\end{equation}

%\begin{equation}
%\mathcal{D}_{t}=\delta_t[\xi^A]+\ou{\omega}{A}{B}(\delta_t)\xi^B
%\end{equation}

\noindent \emph{(ii) bulk and boundary diffeomorphisms.} Next, we consider diffeomorphisms. Dealing with a non-abelian gauge theory, we first lift them from the base manifold into the $SU(2)$ principal bundle (and into its associate vector bundles) over $\mathcal{M}$. This amounts to replacing the ordinary Lie derivative\footnote{The symbol \qq{$\hook$} denotes the interior product $(t\hook\omega)_{a\dots}=t^b\omega_{ba\dots}$ of a $p$-form $\omega_{a\dots}$ with a vector field $t^a$.} $L_t(\cdot)=t\hook\di(\cdot)+\di (t\hook\cdot)$ by the gauge covariant Lie derivative,
\begin{equation}
\begin{split}
&\mathcal{L}_te^i=\nabla(t\hook e^i)+t\hook \nabla e^i,\\
&\mathcal{L}_tq^i=D(t\hook q^i)+t\hook D q^i,
\end{split}\quad
\begin{split}
&\mathcal{L}_tA^i=t\hook F^i,\\
&\mathcal{L}_t\xi^A=t\hook D\xi^A,
\end{split}
\end{equation}
where $D$ denotes the pull-back (see \eref{Ddef}) of the exterior $SU(2)$ gauge covariant derivative $\nabla$ from the bulk to the boundary. These definitions are geometrically meaningful only if the vector field $t^a$ is itself tangential to the boundary, hence
\begin{equation}
t^a\big|_{\mathcal{B}}\in T\mathcal{B}.
\end{equation}
We can now proceed as before. The covariant functional derivative \eref{covD3} of the boundary spinor $\xi^A$ along $\mathcal{L}_t$ is given by
\begin{equation}
\mathcal{D}_t\xi^A=\mathcal{L}_t\xi^A+\frac{1}{2}\tensor{\big(\utilde{\sigma}\mathcal{L}_t\tilde{\sigma}\big)}{^A_B}\xi^B,
\end{equation}
where 
\begin{equation}
\mathcal{L}_t\ou{\tilde{\sigma}}{A}{B}=\ou{\sigma}{A}{Bi}\varphi^\ast_{\mathcal{C}}(\mathcal{L}_tq^i).
\end{equation}
is the Lie derivative of the matrix-valued line density \eref{sigmadens}.

To compute the corresponding charge, consider then a second linearly independent field variation $\delta$ that satisfies the linearised version of the bulk and boundary equations of motion (\ref{fltnss}, \ref{tors}, \ref{glucond}, \ref{bdynmcs}). We contract now both the infinitesimal field variation defined by $\mathcal{L}_t$ and $\delta$ with the pre-symplectic two-form \eref{Omdef}. Taking into account the field equations at the linearised level, e.g.\ $0=\delta F^i=\nabla\delta A^i$, we are left again with a boundary term,
\begin{align}
\nonumber\Omega_\Sigma(\mathcal{D}_t,\delta)=&\frac{1}{8\pi G}\int_\Sigma\Big(\nabla(t\hook e_i)\wedge\delta A^i-\delta e_i\wedge t\hook F^i\Big)+\\
\nonumber&\quad+\frac{1}{2\I}\int_{\mathcal{C}}\Big(\mathcal{L}_t\xi^\dagger_A\ou{\tilde{\sigma}}{A}{B}\delta\xi^B+\frac{1}{2}\xi^\dagger_A\mathcal{L}_t\ou{\tilde{\sigma}}{A}{C}\delta\xi^C+\\
\nonumber&\qquad\quad-\delta\xi^\dagger_A\ou{\tilde\sigma}{A}{B}\delta\xi^B\mathcal{L}_t\xi^B-\frac{1}{2}\delta\xi^\dagger_A\mathcal{L}_t\ou{\tilde\sigma}{A}{B}\xi^B-\CC\Big)=\\
\nonumber=&\frac{1}{2}\int_{\mathcal{C}}\|\xi\|^2(t\hook q_i)\delta A^i+\\
&\quad+\frac{1}{2\I}\int_{\mathcal{C}}\Big(2(\mathcal{L}_t\xi^\dagger_A)\ou{\tilde{\sigma}}{A}{B}\delta\xi^B+\delta\xi^\dagger_A(\mathcal{L}_t\ou{\tilde{\sigma}}{A}{B})\xi^B-\CC\Big).
\end{align}
This boundary term is integrable on phase space for \emph{any} vector field $t^a|_{\mathcal{B}}\in T\mathcal{B}$. This can be seen as follows. First of all, we define the canonical boundary energy momentum tensor density,
\begin{equation}
\uo{\tilde{T}}{i}{a}:=\frac{1}{2\I}\Big(\xi^\dagger_A\ou{\sigma}{A}{Bi}\oepsilon^{ab}D_b\xi^B-\CC\Big),\label{Tdef}
\end{equation}
where $\oepsilon^{ab}$ is the two-dimensional (metric-independent) Levi-Civita tensor density on the boundary $\mathcal{B}=\partial\mathcal{M}$. The tensor density $\uo{\tilde{T}}{i}{a}$ has only tangential components, it is symmetric, traceless (reflecting the conformal invariance of the boundary field theory), and covariantly conserved, i.e.\
\begin{equation}
n^i\uo{\tilde{T}}{i}{a}=0,\quad \uo{\tilde{T}}{i}{a} = q_{ib}q^{ja}\uo{\tilde T}{j}{b},\quad \ou{q}{i}{a}\uo{\tilde{T}}{i}{a}=0,\quad
\ou{q}{j}{b}D_a\uo{\tilde{T}}{j}{a}=0.
\end{equation}
Now, the canonical flux of energy-momentum with respect to an arbitrary vector field $t^a\in T\mathcal{B}$ is given by
\begin{equation}
H_t[\mathcal{C}]=\int_{\mathcal{C}}\ast T_i \ou{q}{i}{a}t^a,\label{diffcharge}
\end{equation}
where $\ast T_i$ is the boundary one-form,
\begin{equation}
T_{ia}=\uo{\tilde{T}}{i}{b}\uepsilon_{ba},
\end{equation}
and $\uepsilon_{ab}$ denotes the inverse density ($\uepsilon_{ab}:\uepsilon_{ab}\oepsilon^{cd}=\delta^c_a\delta^d_b-\delta^c_b\delta^d_a$).

To show that $H_t[\mathcal{C}]$ generates the gauge covariant Lie derivative $\mathcal{L}_t$, we compute the first variation,
\begin{align}\nonumber
\delta H_t[\mathcal{C}]=\frac{1}{2\I}\int_{\mathcal{C}}\Big(\delta\xi^\dagger_A\ou{\sigma}{A}{Bi}t^iD\xi^B+\frac{1}{2\I}\xi^\dagger_A&\ou{\sigma}{A}{Cj}\ou{\sigma}{C}{Bi}\xi^Bt^j\delta A^i+\\
&+\xi^\dagger_A\ou{\sigma}{A}{Bi}t^iD\delta\xi^B-\CC\Big),\label{Hvar}
\end{align}
where $t^i=\ou{q}{i}{a}t^a$ and $\delta\ou{q}{i}{a}=0$, $\delta t^a=0$ on the covariant phase space (the boundary symplectic structure \eref{xipoiss} treats $\ou{\tilde{\sigma}}{A}{B}$ as a fiducial background structure). If we now also take into account the boundary equations of motion \eref{bdynmcs}, we have
\begin{align}
\ou{\sigma}{A}{Bi}t^iD_a\xi^B-&\ou{\sigma}{A}{Bi}\ou{q}{i}{a}\mathcal{L}\xi^B-\frac{1}{2}\ou{\sigma}{A}{Bi}\mathcal{L}_t\ou{q}{i}{a}+\frac{1}{2}D_at^i\ou{\sigma}{A}{Bi}\xi^B=0.
\end{align}
We insert this expression into the variation of the hamiltonian, and immediately find that the functional covariant derivative  is generated by the boundary hamiltonian,
\begin{equation}
\delta H_t[\mathcal{C}]=-\Omega_\Sigma(\mathcal{D}_t,\delta).
\end{equation}
We have thus integrated the Hamilton equation of motion, and shown that $\mathcal{L}_t$ is generated by a hamiltonian, i.e.
\begin{equation}
\mathcal{D}_t(\cdot)=\big\{H_t[\mathcal{C}],\cdot\big\}.
\end{equation}

{Any} generic diffeomorphism\footnote{The  vector field $t^a\in T\mathcal{M}$ is tangential to the two-dimensional cylindrical boundary: $t^a\big|_{\mathcal{B}}\in T\mathcal{B}$.} $\varphi:\mathcal{M}\rightarrow\mathcal{M}, \varphi=\exp(t)$ is, therefore, generated on phase space by a canonical hamiltonian $H_t[\mathcal{C}]$. If, in addition, the vector field $t^a$ induces a conformal Killing vector at the boundary, i.e.\ $\mathcal{L}_tq_{ab}=\eth_ct^c q_{ab}$, then the corresponding charge $H_t[\mathcal{C}]$ will be conserved across the cylinder, i.e.\ $H_t[\mathcal{C}_0]=H_t[\mathcal{C}_1]$. Hence there are infinitely many conserved charges. All of these charges are defined at finite distance\,---\,the boundary $\mathcal{B}$ has a definite topology, but its intrinsic geometry is determined only after having solved the boundary equations of motion \eref{bdynmcs}. In particular, for any regular  solution of the field equations the  circumference of $\mathcal{C}$ will be finite. From the perspective of general relativity in four dimensions, this is a surprise. In four dimensions, expressions for energy and angular momentum exist only asymptotically, and at finite distance a general diffeomorphisms is not integrable (unless particular boundary conditions are imposed, such as those satisfied by spacetimes admitting isolated horizons or Killing horizons).\vspace{0.4em}

\noindent \emph{(iii) length hamiltonian.} Finally, we would like to show that the length of the boundary is a hamiltonian observable as well. Once again, the strategy is to integrate a certain vector field on phase space, and find its hamiltonian generator. Consider thus the following field variation
\begin{equation}
\delta_\alpha\xi^A=-\frac{\alpha}{2\I}(d\ell)^{-1}\ou{\tilde\sigma}{A}{B}\xi^B,
\end{equation}
where $\alpha:\mathcal{C}\rightarrow\R$ is a local gauge parameter and $d\ell$ is the line density
\begin{equation}
d\ell=\sqrt{\tfrac{1}{2}\ou{\tilde\sigma}{A}{B}\ou{\tilde\sigma}{B}{A}}=ds\sqrt{q_{ab}\partial^a_s\partial^b_s}\big|_{\mathcal{C}},\label{linelmnt}
\end{equation}
for a coordinate $s$ on $\mathcal{C}$. The infinitesimal transformation generated by $\delta_\alpha$ only affects the boundary spinor $\xi^A$, all other bulk and boundary fields are constant along $\delta_\alpha$, i.e.\  $\delta_\alpha\ou{e}{i}{a}=\delta_\alpha\ou{A}{i}{a}=\delta_\alpha\ou{q}{i}{a}=0$.

To integrate the vector field $\delta_\alpha$ and find its hamiltonian charge $\delta_\alpha[\cdot]=\{Q_\alpha,\cdot\}$, we proceed as before. We compute the interior product and find
\begin{align}\nonumber
\Omega_\Sigma(\delta_\alpha,\delta)=\frac{1}{2\I}\int_{\mathcal{C}}\Big(\frac{\alpha}{2\I}\xi^\dagger_A(d\ell^{-1})\ou{\tilde\sigma}{A}{C}&\ou{\tilde\sigma}{C}{B}\delta\xi^B+\\
&+\frac{\alpha}{2\I}(d\ell)^{-1}\delta\xi^\dagger_A\ou{\tilde\sigma}{A}{C}\ou{\tilde\sigma}{C}{B}\xi^A-\CC\Big).\label{lvar}
\end{align}
The square of the matrix-valued densities $\ou{\tilde{\sigma}}{A}{B}$ is proportional to the identity,
\begin{equation}
\ou{\tilde{\sigma}}{A}{C}\ou{\tilde{\sigma}}{C}{B}=(d\ell)^2\delta^A_B.
\end{equation}
We thus have a total derivative on phase space\footnote{On phase space, the boundary frame field $\ou{q}{i}{a}$ is treated as a fiducial background structure, hence $\delta[\ou{q}{i}{a}]=0$ and $\delta [d\ell]=0$.}
\begin{equation}
\Omega_\Sigma(\delta_\alpha,\delta)=-\frac{1}{2}\int_{\mathcal{C}}d\ell\,\alpha\,\delta\|\xi\|^2=-\delta Q_\alpha[\mathcal{C}].
\end{equation}
Now, the $SU(2)$ norm $\|\xi\|^2$ of the boundary spinor is nothing but the conformal factor that relates the unphysical boundary metric to the pull-back of the three-dimensional space-time metric, namely
\begin{equation}
\varphi^\ast_{\mathcal{B}} g_{ab}=(4\pi G)^2\|\xi\|^4q_{ab}.
\end{equation}
We can express, therefore, the canonical charge in terms of the physical metric alone, hence
\begin{equation}
Q_\alpha[\mathcal{C}]=\frac{1}{2}\int_{\mathcal{C}}d\ell\,\alpha\|\xi\|^2=\frac{1}{8\pi G}\int_{\mathcal{C}}\alpha\sqrt{e_ie^i}.
\end{equation}
The length of our one-dimensional boundary is  
the zero mode of this observable,
\begin{equation}
\text{Length}[\mathcal{C}]=8\pi G\,Q_1[\mathcal{C}].\label{Lcharge}
\end{equation}

\subsection{Physical phase space and quantisation of length}\noindent
We now want to explore some aspects of the resulting quantum theory. The first step is to compute the pull-back of the pre-symplectic potential \eref{thetadef} to the physical phase space and identify the canonical coordinates. 

In the interior, the curvature vanishes, hence the connection is pure gauge
\begin{equation}
A_a=g^{-1}\partial_ag,
\end{equation}
where $\partial_a$ is some flat reference connection. The functional differential of this connection gives the covariant derivative of the Maurer\,--\,Cartan form,
\begin{equation}
\bbvar{d}A_a=\nabla_a(g^{-1}\bbvar{d}g).
\end{equation}

Next, we decompose the boundary spinor into the eigen-spinors of the matrix-valued line element ${\ou{\tilde{\sigma}}{A}{B}}$ that enters the Poisson brackets \eref{xipoiss} at the pre-symplectic level. We thus write,
\begin{equation}
\xi^A=(d\ell)^{-\frac{1}{2}}\big(\bar{a}o^A+{b}\iota^A\big),
\end{equation}
where $d\ell$ is the fiducial line element \eref{linelmnt} and the normalised eigen-spinors $o^A$ and $\iota^A$ satisfy 
\begin{subalign}
\ou{\tilde{\sigma}}{A}{B}o^B&=+(d\ell)o^A,\\
\ou{\tilde{\sigma}}{A}{B}\iota^B&=-(d\ell)\iota^A.
\end{subalign}
Notice that the components $a$ and $b$ are half-densities on $\mathcal{C}$.

We can now evaluate the symplectic potential. Taking into account the torsionless condition \eref{tors} in the bulk, we are left with a boundary integral along the perimeter $\mathcal{C}=\partial\Sigma$ of the disk,
\begin{equation}
\Theta_\Sigma=-\frac{1}{4\pi G}\int_{\mathcal{C}}\tilde{e}^i\mathrm{Tr}(\tau_ig^{-1}\bbvar{d}g)+\frac{1}{2\I}\int_{\mathcal{C}}\big(a\bbvar{d}\bar{a}-\bar{b}\bbvar{d}b-\CC\big),
\end{equation}
where $\tau_i=(2\I)^{-1}\sigma_i$ are the $\mathfrak{su}(2)$ generators and $\tilde{e}^i$ is the vector-valued line density
\begin{equation}
\tilde{e}^i=\varphi^\ast_{\mathcal{C}}e^i.
\end{equation}
The only non-vanishing Poisson brackets derived from this symplectic potential are given by
\begin{subalign}
\big\{\tilde{e}_i(s),\ou{g}{A}{B}(s')\big\}&=-8\pi G\,\tilde{\delta}^{(1)}(s,s')\,\ou{\big[g\tau_i\big]}{A}{B}(s),\\
\big\{\tilde{e}_i(s),\tilde{e}_j(s')\big\}&=-8\pi G\,\tilde{\delta}^{(1)}(s,s')\uo{\epsilon}{ij}{k}\tilde{e}_k(s),\\
\big\{a(s),\bar{a}(s')\big\}&=\big\{b(s),\bar{b}(s')\big\}=\I\tilde{\delta}^{(1)}(s,s').\label{harmosci}
\end{subalign}

This is not yet the physical phase space. We still have to impose the glueing conditions \eref{glucond}, namely
\begin{subalign}
\tilde{c}&=\tilde{e}_i\ell^i-4\pi G(\bar{a}a+\bar{b}b)=0,\label{fclass}\\
\tilde{c}_{\pm}&=\tilde{e}_in^i\pm\I\tilde{e}_i\ou{\varepsilon}{i}{j}\ell^j=0,
\end{subalign}
where $n^i$ is the internal normal vector \eref{bndryn}, and $\ell^i:=\ou{q}{i}{a}\ell^a$ is a normalised tangent vector to the boundary $\mathcal{C}$: $\ell^a\in T\mathcal{C}:q_{ab}\ell^a\ell^b=1$, whose direction follows  the orientation of $\mathcal{C}$. The triple $(n^i,\ou{\varepsilon}{i}{k}\ell^k,\ell^i)$ defines, therefore, a positively oriented triad.

The constraint $\tilde{c}=0$ is first class. The constraints $\tilde{c}_{\pm}=0$, on the other hand, are second class. The resulting Dirac bracket is
\begin{align}
\big\{A,B\big\}^\ast&=\big\{A,B\big\}-\frac{\I}{16\pi G}\int_{\mathcal{C}}\utilde{e}\Big[\big\{A,\tilde{c}_+\big\}\big\{\tilde{c}_-,B\big\}-(A\leftrightarrow B)\Big],
\end{align}
where $\utilde{e}$ is the inverse line density: $\utilde{e}=(\tilde{e}_i\ell^i)^{-1}$.

The crucial point is now that the Dirac bracket leaves the commutation relations for the oscillators \eref{harmosci} untouched, 
\begin{equation}
\big\{a(s),\bar{a}(s')\big\}^\ast=\big\{b(s),\bar{b}(s')\big\}^\ast=\I\tilde{\delta}^{(1)}(s,s').
\end{equation}
The components of the gauge element, on the other hand, turn out to be Poisson non-commutative,
\begin{equation}
\begin{split}
\big\{\ou{g}{A}{B}(s),\ou{g}{C}{D}(s')\big\}^\ast=&-4\pi\I G\,\tilde{\delta}^{(1)}(s,s')\ou{\big[g\tau_+\big]}{A}{B}\utilde{e}\ou{\big[g\tau_-\big]}{C}{D}+\\
&+4\pi\I G\,\tilde{\delta}^{(1)}(s,s')\ou{\big[g\tau_-\big]}{A}{B}\utilde{e}\ou{\big[g\tau_+\big]}{C}{D},
\end{split}
\end{equation}
where $\tau_\pm=\tau_in^i\pm\I\tau_i\ou{\varepsilon}{i}{j}\ell^j$.\vspace{0.4em}

In quantum theory, the oscillators $a$, $\bar{a}$ and $b$ and $\bar{b}$, turn into creation and annihilation operators. Assuming bosonic commutation relations, we thus have for all $s,s'\in\mathcal{C}$,\ that
\begin{equation}
\big[a(s),a^\dagger(s')\big]=\big[b(s),b^\dagger(s')\big]=\tilde{\delta}(s,s').
\end{equation}
The analogue of the Ashtekar\,--\,Lewandowski vacuum \cite{Ashtekar:1994mh} in the continuum is then given by the state $|\emptyset\rangle_{\mathcal{C}}$ that is annihilated by $a$ and $b$, namely
\begin{equation}
a(s)\big|\emptyset\big\rangle_{\mathcal{C}}=b(s)\big|\emptyset\big\rangle_{\mathcal{C}}=0.
\end{equation}

In \eref{Lcharge}, we saw that the total length of the of the one-dimensional boundary is given by the line integral
\begin{equation}
\mathrm{Length}[\mathcal{C}]=\int_{\mathcal{C}}\sqrt{e_ie^i}=\int_{\mathcal{C}}\tilde{e}_i\ell^i.
\end{equation}
The glueing condition \eref{fclass}, which is first class, tells us that this observable is proportional to the norm $\|\xi\|^2=\bar{a}a+b\bar{b}$ of the boundary spinor. Choosing a normal ordering, we have
\begin{equation}
\boldsymbol{\colon}\!\mathrm{Length}[\mathcal{C}]\boldsymbol{\colon}=4\pi G\int_{\mathcal{C}}\big(a^\dagger a+b^\dagger b\big).
\end{equation}
In quantum theory, this is nothing but the number operator for the two oscillators. This operator has a discrete spectrum. A hypothetical observer %in our three-dimensional Euclidean world-tube 
that sets up an experiment and measures the area of a one-dimensional cross-section of the boundary, will see, therefore, only the following measurement outcomes
\begin{equation}
\ell_n=4\pi G\, n,\quad n=0,1,2,\dots.\label{Lspectrm}
\end{equation}
%A similar argument applies in four dimensions for Lorentzian signature \cite{Wieland:2017cmf}.
%\section*{\color{red}{test}}
%\begin{equation}
%%\rotatebox{180}{{\text{\small$\,\bbgreek{\Gamma}\,$}}}
%\end{equation}

\section{Conclusion}
\noindent Let me summarise. In this paper, we studied euclidean gravity in an infinite cylinder of finite radius (see \hyperref[fig1]{figure 1}). At the boundary of this cylinder, a counter term is required to cancel the boundary term appearing in the first variation of the action. We then introduced such a boundary term by coupling an $SU(2)$ boundary spinor to the $SU(2)$ spin connection in the bulk. Having introduced a boundary action, we deduced the corresponding boundary equations of motion: the glueing condition \eref{glucond} implies that the conformal factor between the fiducial boundary metric $q_{ab}=\delta_{ij}\ou{q}{i}{a}\ou{q}{j}{b}$ and the physical metric $g_{ab}$ in the bulk is determined by the $SU(2)$ norm of the boundary spinor. We also found the equations of motion for the boundary spinor itself and identified them with a two-dimensional analogue of the Witten equations that appear in the celebrated proof of the positive mass theorem \cite{Wittenproof}. 

%:
Next, we studied the canonical formulation and the underlying gauge symmetries in the covariant hamiltonian formalism \cite{Wald:1999wa}. In deriving the Hamilton equations of motion from the variation of the action a subtlety arises, because the boundary symplectic structure may be itself time dependent: the Poisson brackets \eref{xipoiss} for the boundary spinor $\xi^A$ are determined by the densitised Pauli matrix $\ou{\tilde{\sigma}}{A}{B}=\varphi^\ast_{\Sigma_{t}}\ou{\sigma}{A}{B}$, and for a general foliation and a general vector field $t^a\in T\mathcal{B}$, this Pauli matrix will be time-dependent itself: $\mathcal{L}_t\ou{\tilde{\sigma}}{A}{B}\neq 0$. This implicit time dependence then modifies the left hand side of the Hamilton equations: the ordinary Lie derivative is replaced by a covariant derivative with respect to a connection on phase space.\footnote{The corresponding curvature $\ou{\mathcal{F}}{A}{B}(\delta_t,\delta_{t'})\xi^B=\mathcal{D}_t\mathcal{D}_{t'}\xi^A-\mathcal{D}_{t'}\mathcal{D}_{t}\xi^A-\mathcal{D}_{[t,t']}\xi^A$ is a measure for the central charge $K(t,t')=\{H_t,\{H_{t'},\cdot\}\}-\{H_{t'},\{H_{t},\cdot\}\}-\{H_{[t,t']},\cdot\}$. The recent paper of Aldo Riello and Henrique Gomes investigate the geometry of such field space connections on a more general level \cite{Gomes:2016mwl}.} Gauge symmetries are then given by the degenerate direction of the pre-symplectic two-form \eref{Omdef}. Internal $SU(2)$ gauge transformations and small diffeomorphisms in the bulk are indeed gauge symmetries of the theory. On the other hand, large diffeomorphisms $\varphi:\mathcal{M}\rightarrow\mathcal{M}$ that preserve the boundary $\varphi(\partial\mathcal{M})=\partial\mathcal{M}$ do not annihilate the pre-symplectic two-form (hence do not define gauge symmetries) but are generated by a hamiltonian, whose on-shell value is given by the canonical energy-momentum flux with respect to the energy-momentum tensor \eref{Tdef} of the boundary field theory.\footnote{An altogether different treatment for these diffeomorphism charges is being developed by Freidel, Donnelly \cite{Donnelly:2016auv} and Geiller \cite{Geiller:2017xad}. In this new extended phase space approach the gauge parameters ($\mathfrak{su}(2)$ Lie algebra elements for internal gauge transformations, vector fields for diffeomorphisms) are considered as new canonical variables that are added to the symplectic structure to restore gauge invariance at the boundary.} 

Finally, we discussed some aspects of the resulting quantum theory. First of all, we considered the symplectic structure on the physical phase space. The corresponding symplectic potential is an integral over a one-dimensional cross section $\mathcal{C}$ of the boundary $\mathcal{B}=\partial\mathcal{M}$ (see \hyperref[fig1]{figure 1}). The canonical variables are given by an $SU(2)$ gauge element $\ou{g}{A}{B}:\mathcal{C}\rightarrow SU(2)$, a conjugate Lie algebra-valued line density $\tilde{e}^i$ and the spin up and down components ($\bar{a}$ and $b$) of the boundary spinor $\xi^A$. The glueing conditions are constraints on this phase space. Two of them are second class, the other one is a first class constraint generating a $U(1)$ gauge symmetry. Introducing the Dirac bracket on the physical phase space, we then saw that $a$, $b$ and $\bar{a}$, $\bar{b}$ satisfy the canonical Poisson commutation relations of the harmonic oscillator.  The conformal factor is an observable on this phase space. It is simply given by the number operator that counts the number of quanta excited over the Fock vacuum $|\emptyset\rangle_{\mathcal{C}}$, which is the state annihilated by $a$, $b$ and $\tilde{e}^i$,
\begin{equation}
\forall s\in\mathcal{C}:a(s)|\emptyset\rangle_{\mathcal{C}}=b(s)|\emptyset\rangle_{\mathcal{C}}=\tilde{e}^i(s)|\emptyset\rangle_{\mathcal{C}}=0.\label{Fvacu}
\end{equation}
The quantisation of the conformal factor immediately implies then the quantisation of length: the total circumference of the boundary is quantised, and all possible eigenvalues of length \eref{Lspectrm} are a multiple of the fundamental Planck length. This result is well known from three-dimensional loop quantum gravity \cite{Rovelli:1993kc,Freidel:2002hx,Thiemann:1997ru}, but in here it is derived using a very different representation of the canonical commutation relations. In fact, spin networks and triangulations of space never entered the construction.

The Fock vacuum \eref{Fvacu} represents a configuration where the boundary $\mathcal{C}$ has shrunken to a point. It is the ground state of geometry, but not of energy itself. This is in complete analogy with the Ashtekar\,--\,Lewandowski vacuum in four dimensions: the Ashtekar\,--\,Lewandowski vacuum \cite{Ashtekar:1994mh} describes a totally degenerate three-geometry, it is the ground state of geometry, but it is not the state of minimal energy (in asymptotically flat spacetimes, this would be the Minkowski vacuum). The same happens here: the \emph{geometry-vacuum} $|\emptyset\rangle_{\mathcal{C}}$ is not the \emph{energy-vacuum} of the theory. In fact, the canonical energy  \eref{diffcharge} for a vector field $t^a\in T\mathcal{C}^\perp$, $q_{ab}t^at^b=1$  will only vanish on average,
\begin{equation}
H_t[\mathcal{C}]|\emptyset\tensor{\rangle}{_{\mathcal{C}}}\neq 0,\quad\text{but}\quad\tensor[_{\mathcal{C}}]{\langle}{}\emptyset|H_t[\mathcal{C}]|\emptyset\tensor{\rangle}{_{\mathcal{C}}}=0.
\end{equation}
This can be seen by writing $H_t[\mathcal{C}]$ in terms of the harmonic oscillators, which yields a squeeze operator $H_t[\mathcal{C}]\sim\I\oint_{\mathcal{C}}(a^\dagger Db^\dagger-\HC)$. It is therefore not very surprising that the geometry-vacuum $|\emptyset\tensor{\rangle}{_{\mathcal{C}}}$ is not an eigenstate of energy. Moreover, for any possible cross section $\mathcal{C}, \mathcal{C}',\dots$ there will be a different such vacuum $|\emptyset\tensor{\rangle}{_{\mathcal{C}}}, |\emptyset\tensor{\rangle}{_{\mathcal{C}'}},\dots$  Recent developments have stressed the potential significance of such vacuum ambiguities for the black hole information paradox \cite{Hawking:2016msc}. The results from this paper suggest that these vacuum ambiguities appear in the loop quantum gravity continuum limit as well, and amount to choosing different cross sections of the boundary.
\vspace{0.4em}

\noindent\emph{Acknowledgments.} I gratefully acknowledge that this research was supported from Perimeter Institute for Theoretical Physics. Research at Perimeter Institute is supported by the Government of Canada through the Department of Innovation, Science and Economic Development and by the Province of Ontario through the Ministry of Research and Innovation.

\section*{Appendix: $SU(2)$ Spinors}\label{appndx}
\noindent The spin connection $\ou{A}{i}{a}$ acts naturally on the associated spin bundle through the gauge covariant derivative
\begin{equation}
\nabla_a\xi^A=\partial_a\xi^A+\frac{1}{2\I}\ou{A}{i}{a}\ou{\sigma}{A}{Bi}\xi^B,
\end{equation}
where $\partial_a$ is a derivative for a flat reference connection and $\ou{\sigma}{A}{Bi}$ are the three-dimensional Pauli matrices that satisfy the familiar Pauli identity
\begin{equation}
\ou{\sigma}{A}{Ci}\ou{\sigma}{C}{Bj}=\delta^A_B\delta_{ij}+\I\uo{\epsilon}{ij}{k}\ou{\sigma}{A}{Bk},\label{Pauliid}
\end{equation}
where $A,B,C,\dots$ are abstract spinor indices. Since $SU(2)$ is unitary, there exists an $SU(2)$ invariant hermitian metric\footnote{Indices $A',B',C',\dots$ refer to the complex conjugate representation.}
\begin{equation}
\langle \phi|\psi\rangle=\delta_{AA'}\bar{\phi}^{A'}\psi^A,\quad\|\xi\|^2=\delta_{AA'}\bar{\xi}^{A'}\xi^A,
\end{equation}
which we use to define the conjugate spinors
\begin{equation}
\phi^\dagger_A:=\delta_{AA'}\bar{\phi}^{A'}.
\end{equation}
Both the Pauli matrices (the internal soldering forms) as well as the internal metric are all annihilated by the connection, in other words $\nabla_a\delta_{AA'}=0$ and $\nabla_a\ou{\sigma}{A}{Bi}=0$.
%\section*{References}
\providecommand{\href}[2]{#2}\begingroup\raggedright\endgroup

\end{document}